\documentclass[twocolumn,showpacs,preprintnumbers,amsmath,amssymb,epsf]{revtex4}
\usepackage{graphicx}
\usepackage{dcolumn}
\usepackage{bm}

\begin{document}

preprint{APS/123-QED}

\title{Inflow versus outflow zero-temperature dynamics in one dimension.}

\author{Katarzyna Sznajd--Weron and Sylwia Krupa}
\email{kweron@ift.uni.wroc.pl}
\homepage{http://www.ift.uni.wroc.pl/~kweron}
\affiliation{Institute of Theoretical Physics, University of
Wroc{\l}aw, pl. Maxa Borna 9, 50-204 Wroc{\l}aw, Poland }

\date{\today}

\begin{abstract}
It has been suggested that Glauber (inflow) and Sznajd (outflow) zero-temperature dynamics for the one dimensional Ising ferromagnet with the nearest neighbors interactions are equivalent. Here we compare both dynamics from analytical and computational points of view. We use the method of mapping an Ising spin system onto the dimer RSA model and show that already this simple mapping allows to see the differences between inflow and outflow zero-temperature dynamics. Then we investigate both dynamics with synchronous, partially synchronous and random sequential updating using the Monte Carlo technique and compare both dynamics in terms of the number of persistent spins, clusters, mean relaxation time and relaxation time distribution. 
\end{abstract}

\pacs{05.50.+q}%
\maketitle

\section{Introduction}
The majority of natural phenomena observed in physics, biology, geology, social sciences, etc., are non-equilibrium processes. Unfortunately, the theory of non-equilibrium statistical mechanics is far less developed than its equilibrium part. As a result, the most ubiquitous phenomena are poorly understood \cite{SZ98}.
The zero-temperature dynamics of simple models, such as Ising ferromagnets, provides interesting examples of non-equilibrium dynamical systems with many attractors (absorbing configurations, blocked configurations, zero-temperature metastable states) \cite{GL04}. In this paper we focus on the so-called single spin flip dynamics for the one dimensional Ising ferromagnet. The best known example of such a dynamics for the Ising model is Glauber dynamics \cite{Glauber}. It can be viewed as "inflow" dynamics, since the center spin is influenced by its nearest neighbors \cite{Krupa}. Another type of dynamics, which can be called "outflow" dynamics, since the information flows from the center spin (or spins) to the neighborhood, has been introduced to describe opinion formation in social systems \cite{SWS00}. It has been suggested \cite{g05, bs03} that both dynamics for the Ising ferromagnet with the nearest neighbors interactions are equivalent, at least in one dimension. However, it seems to be true only in some particular cases. The aim of this paper is to compare generalized outflow and inflow dynamics for a chain of Ising spins and show in which cases there are equivalent and in which they differ. 

In the first two sections we recall ideas of inflow and outflow dynamics and formulate the generalized versions of both dynamics. We take both dynamics under a common roof reformulating them without using the concept of energy. In the third section we use the illuminating method of mapping an Ising spin system onto the dimer RSA model and make simple mean field like calculations to show the difference between the two dynamics. In the fourth section we present Monte Carlo results for both dynamics with several kinds of updating, including synchronous, partially synchronous and random sequential updating. The summary of simulation results and conclusions are the subjects of the fifth and sixth section of the paper.

\section{Inflow dynamics}
The best known example of such a dynamics for the Ising model is Glauber dynamics. Within Glauber dynamics, in a broad sense, each spin is flipped $S_i(\tau) \rightarrow -S_i(\tau+1)$ with a rate $W(\delta E)$ per unit time and this rate is assumed to depend only on the energy difference implied by the flip \cite{GL04}. 
The two most common choices of flipping rates in the case of discrete updates are heat-bath and Metropolis, both obey the detailed balance condition:
\begin{equation}
\frac{W(\delta E)}{W(-\delta E)}= \exp (-\beta \delta E).
\label{db}
\end{equation}

Recently it was shown \cite{GL04} that there was a vast family of dynamical rates, besides these two choices, which obeyed condition (\ref{db}). Among them the class of zero-temperature dynamics defined as:
\begin{eqnarray}
W(\delta E)=
\left\{
\begin{array}{ll}
1 & \mbox{if }  \delta E<0, \\ 
W_0 & \mbox{if }  \delta E=0, \\ 
0 & \mbox{if }  \delta E>0. 
\end{array}
\right.
\label{w}
\end{eqnarray}
occurred to be very interesting. The zero-temperature limits of the heat-bath and Metropolis rates are  $W^{HB}_0 = 1/2$ and $W^M_0 = 1$, respectively. 
For any non-zero value of the rate $W_0$ corresponding to free spins, the dynamics belongs to the universality class of the zero-temperature Glauber model. This is a prototypical example of phase ordering
by domain growth (coarsening). The typical size of ordered domains of consecutive
$\uparrow$ and $\downarrow$ spins grows as $L(t) \sim t^{1/2}$. The particular value $W_0 = 0$ corresponds to the constrained zero-temperature Glauber dynamics (\cite{GL04} and references therein).
In the constrained zero-temperature Glauber dynamics, the only possible moves are flips
of isolated spins and the system therefore eventually reaches a blocked configuration, where there is no isolated spin \cite{GL04}. Very interesting results for the zero-temperature Glauber dynamics have been also obtained using computer simulations \cite{RK98,SKR01,SKR02,lipowski}.

In out of equilibrium systems, there is usually no energy function and the system is only defined by its dynamical rules \cite{D98}. This is also the case of the sociophysics Sznajd model. For this reason we reformulate the definition of the zero-temperature Glauber dynamics for the Ising ferromagnet, without using the concept of the energy, in the following way:
\begin{eqnarray}
\begin{array}{l}
S_i(\tau+1)  = \\
\left\{
\begin{array}{ll}
1 & \mbox{if }  \sum_{nn} S_{nn}>0, \\ 
-S_i(\tau) \hspace{0.2cm} \mbox{with prob} \hspace{0.2cm} W_0 & \mbox{if } \sum_{nn} S_{nn}=0, \\ 
-1 & \mbox{if }  \sum_{nn} S_{nn}<0, 
\end{array}
\right. 
\end{array}
\end{eqnarray}
where $\sum_{nn} S_{nn}$ denotes the sum over nearest neighbors. 

In one dimension, which is the case of this paper, the above definition can be written as:
\begin{eqnarray}
\begin{array}{l}
S_i(\tau+1)= \\
\left\{
\begin{array}{ll}
1 & \mbox{if }  S(\tau)_{i-1}+S(\tau)_{i+1}>0, \\ 
-S_i(\tau) \hspace{0.2cm} \mbox{with prob} \hspace{0.2cm} W_0 & \mbox{if } S(\tau)_{i-1}+S(\tau)_{i+1}=0, \\ 
-1 & \mbox{if }  S(\tau)_{i-1}+S(\tau)_{i+1}<0. 
\end{array}
\right.
\end{array}
\label{in}
\end{eqnarray}

\section{Outflow dynamics}

The outflow dynamics was introduced to describe opinion changes in the society. The idea was based on the fundamental social phenomenon called "social validation". However, in this paper we do not focus on social applications of the model (interesting reviews can be found in \cite{Stauffer02,Schechter02,FS05,SW05}). On the contrary, here we investigate the dynamics from the theoretical point of view.  

In the original model a pair of neighboring spins $S_i$ and $S_{i+1}$ is chosen and if $S_iS_{i+1}=1$ the the two neighbors of the pair follow its direction, i.e. $S_{i-1} \rightarrow S_{i}(=S_{i+1})$ and $S_{i+2} \rightarrow S_{i}(=S_{i+1})$.
Such a rule has been used also in all later papers dealing with the one dimensional case of the model. However, the case in which $S_iS_{i+1}=-1$ is far less obvious. For example, in the original paper  $S_iS_{i+1}=-1$ led to $S_{i-1} \rightarrow S_{i+1}$ and $S_{i+2} \rightarrow S_{i}$. However, some authors have argued \cite{Sanchez04} that such a rule is unrealistic in a model trying to represent the behavior of a community. Moreover, the original Sznajd model with both the ferromagnetic and the antiferomagnetic rules is equivalent to a single simple rule that every spin takes the direction of its next nearest neighbor independently of the product $S_iS_{i+1}$. To see this observe that the ferromagnetic rule:\\
If $S_i(\tau)S_{i+1}(\tau)=1$ then 
$S_{i-1}(\tau+1) \rightarrow S_{i}(\tau)$ and $S_{i+2}(\tau+1) \rightarrow S_{i+1}(\tau)$ \\
is equivalent to the rule: \\ if $S_i(\tau)S_{i+1}(\tau)=1$ then  $S_{i-1}(\tau+1) \rightarrow S_{i+1}(\tau)$ i $S_{i+2}(\tau+1) \rightarrow S_{i}(\tau)$; \\ while the antiferromagnetic rule states that\\
if $S_i(\tau)S_{i+1}(\tau)=-1$ then 
$S_{i-1}(\tau+1) \rightarrow S_{i+1}(\tau)$ and $S_{i+2}(\tau+1) \rightarrow S_{i}(\tau)$.

Thus, the two above rules can be rewritten as a simple single rule:
$S_{i-1}(\tau+1) \rightarrow S_{i+1}(\tau)$ and $S_{i+2}(\tau+1) \rightarrow S_{i}(\tau)$. 

In later papers \cite{SWW03,SWW02} we have proposed two modifications of the model in which the antiferromagnetic rule was replaced by one of rules described below:  
\begin{description}
\item[modification 1:] If $S_i(\tau)S_{i+1}(\tau)=-1$:
$S_{i-1}(\tau+1) \rightarrow S_{i-1}(\tau)$ and $S_{i+2}(\tau+1) \rightarrow S_{i+2}(\tau)$. 
\end{description}
or
\begin{description}
\item[modification 2:] If $S_i(\tau)S_{i+1}(\tau)=-1$:
$S_{i-1}(\tau+1) \rightarrow -S_{i-1}(\tau)$ and $S_{i+2}(\tau+1) \rightarrow -S_{i+2}(\tau)$ with probability 1/2. 
\end{description}

A generalized dynamics, which includes the two above modifications, can be written as:
\begin{eqnarray}
\begin{array}{l}
S_i(\tau+1)= \\
\left\{
\begin{array}{ll}
1 & \mbox{if }  S_{i+1}(\tau)+S_{i+2}(\tau)>0, \\ 
-S_i(\tau) \hspace{0.2cm} \mbox{with prob} \hspace{0.2cm} W_0 & \mbox{if } S_{i+1}(\tau)+S_{i+2}(\tau)=0, \\ 
-1 & \mbox{if }  S_{i+1}(\tau)+S_{i+2}(\tau)<0. 
\end{array}
\right.
\end{array}
\label{out}
\end{eqnarray}

It is easy to notice that modification 1 corresponds to $W^1_0=0$ and modification 2 to $W^2_0=1/2$.

\section{Mapping onto the dimer model}
As was mentioned in the previous sections, for $W_0=0$ the system under inflow (Glauber) dynamics described by formula (\ref{in}) eventually reaches a blocked configuration, where there is no isolated spin. On the other hand, the system under the outflow dynamics described by (\ref{out}) always reaches the ferromagnetic steady state. Thus, for $W_0=0$ the difference between the outflow and inflow dynamics is obvious. Nevertheless, within the mean field approach \cite{SL03} and the Galam's unifying frame \cite{g05} both dynamics are equivalent, i.e. there is no difference between the outflow and inflow dynamics, even for $W_0=0$

Here we use the illuminating method of mapping the Ising spin system onto the dimer RSA model. This has been already done for the inflow dynamics \cite{GL04}:
\begin{equation}
\left\{
\begin{array}{lll}
X_i=S_{i}S_{i+1}=1 & \Rightarrow & \circ, \\ 
X_i=S_{i}S_{i+1}=-1 & \Rightarrow & \bullet.
\end{array}
\right.
\end{equation}

In the case of inflow dynamics the following transitions, which change the state of the system, are possible:

\begin{center}
\begin{tabular}{c|c}
spins & particles \\
\hline
\hline
$\downarrow \uparrow \downarrow \rightarrow  \downarrow \downarrow \downarrow$ & $\bullet\bullet \rightarrow \circ\circ$ \\ 
$\uparrow \downarrow \uparrow \rightarrow  \uparrow \uparrow \uparrow$ & $\bullet\bullet \rightarrow \circ\circ$ \\ 
\hline
$\downarrow \uparrow \uparrow \stackrel{W_0}{\rightarrow}  \downarrow \downarrow \uparrow$ & $\bullet\circ \stackrel{W_0}{\rightarrow} \circ\bullet$ \\ 
$\uparrow \downarrow \downarrow \stackrel{W_0}{\rightarrow}  \uparrow \uparrow \downarrow$ & $\bullet\circ \stackrel{W_0}{\rightarrow} \circ\bullet$ \\
\hline 
$\downarrow \downarrow \uparrow \stackrel{W_0}{\rightarrow}  \downarrow \uparrow \uparrow$ & $\circ\bullet \stackrel{W_0}{\rightarrow} \bullet\circ$ \\ 
$\uparrow \uparrow \downarrow \stackrel{W_0}{\rightarrow}  \uparrow \downarrow \downarrow$ & $\circ\bullet \stackrel{W_0}{\rightarrow} \bullet\circ$ \\
\end{tabular}
\end{center}
Thus, after the mapping there are only two types of transitions for the inflow dynamics: $\bullet\bullet \rightarrow \circ\circ$ and $\circ\bullet \stackrel{W_0}{\leftrightarrow} \bullet\circ$.
This mapping shows at once that for $W_0=0$ the dynamics is fully irreversible, in the sense that
each spin flips at most once during the whole history of the sample.

It should be noticed that if we map the system under the outflow dynamics onto the dimer model we have to take into account four particles, because changing the border spin influences the next particle. In this
case four types of transitions are possible: $\circ\bullet\circ \rightarrow \circ\circ\bullet$,
$\circ\bullet\bullet \rightarrow \circ\circ\circ$, $\bullet\circ\bullet \stackrel{W_0}{\leftrightarrow} \bullet \bullet \circ$ and $\bullet\bullet\bullet \stackrel{W_0}{\leftrightarrow} \bullet\circ\circ$ (to make it clearer the flipped spins are denoted by double arrows in the table below):
\begin{center}
\begin{tabular}{c|c}
spins & particles \\
\hline
\hline
$\downarrow \downarrow \Uparrow \uparrow \rightarrow  \downarrow \downarrow \Downarrow \uparrow$ & $\circ\bullet\circ \rightarrow \circ\circ\bullet$ \\ 
$\uparrow \uparrow \Downarrow \downarrow \rightarrow  \uparrow \uparrow \Uparrow \downarrow$ & $\circ\bullet\circ \rightarrow \circ\circ\bullet$\\ 
\hline
$\downarrow \downarrow \Uparrow  \downarrow \rightarrow  \downarrow \downarrow \Downarrow \downarrow$ & $\circ\bullet\bullet \rightarrow \circ\circ\circ$ \\ 
$\uparrow \uparrow \Downarrow \uparrow \rightarrow  \uparrow \uparrow \Uparrow \uparrow$ & $\circ\bullet\bullet \rightarrow \circ\circ\circ$ \\ 
\hline
$\downarrow \uparrow \Uparrow \downarrow \stackrel{W_0}{\rightarrow} \downarrow \uparrow \Downarrow \downarrow$ & $\bullet\circ\bullet \stackrel{W_0}{\rightarrow} \bullet \bullet \circ$\\
$\uparrow \downarrow \Downarrow \uparrow \stackrel{W_0}{\rightarrow} \uparrow \downarrow \Uparrow \uparrow$ & $ \bullet \circ \bullet \stackrel{W_0}{\rightarrow} \bullet\bullet\circ$\\
\hline
$\uparrow \downarrow \Uparrow \downarrow \stackrel{W_0}{\rightarrow} \uparrow \downarrow \Downarrow \downarrow$ & $\bullet\bullet\bullet \stackrel{W_0}{\rightarrow} \bullet\circ\circ$\\
$\downarrow \uparrow \Downarrow \uparrow \stackrel{W_0}{\rightarrow} \downarrow \uparrow \Uparrow \uparrow$ & $\bullet\bullet\bullet \stackrel{W_0}{\rightarrow} \bullet\circ\circ$\\
\hline
$\downarrow \uparrow \Downarrow \downarrow \stackrel{W_0}{\rightarrow} \downarrow \uparrow \Uparrow \downarrow$ & $\bullet\bullet\circ \stackrel{W_0}{\rightarrow} \bullet\circ\bullet $\\
$\uparrow \downarrow \Uparrow \uparrow \stackrel{W_0}{\rightarrow} \uparrow \downarrow \Downarrow \uparrow$ & $\bullet\bullet\circ \stackrel{W_0}{\rightarrow} \bullet \circ \bullet$\\
\hline
$\uparrow \downarrow \Downarrow \downarrow \stackrel{W_0}{\rightarrow} \uparrow \downarrow \Uparrow \downarrow$ & $\bullet\circ\circ \stackrel{W_0}{\rightarrow} \bullet\bullet\bullet$\\
$\downarrow \uparrow \Uparrow \uparrow \stackrel{W_0}{\rightarrow} \downarrow \uparrow \Downarrow \uparrow$ & $\bullet\circ\circ \stackrel{W_0}{\rightarrow} \bullet \bullet \bullet$\\
\end{tabular}
\end{center}

This mapping shows that for $W_0=0$ the outflow dynamics consists of two processes - diffusion of $\bullet$ particles in the sea of $\circ$ empty sites and annihilation of $\bullet \bullet$ pairs. Thus our model for $W_0$ with random sequential updating reduces to an analytically solvable reaction-diffusion system $A+A \rightarrow 0$ (denote by $\circ$ the empty place and $\bullet$ by the $A$ particle).

For $W_0 \ge 0$ we can also use the mean field approach (MFA). The mean field results for the outflow dynamics without dimer mapping can be found in \cite{SL03}.
Within dimer mapping we take into account correlations between pairs of the nearest neighbors. Thus if we apply MFA to the mapped system we can expect more correct results than obtained and within MFA without mapping. 

Let us denote the number of $\bullet$ particles by $N_b$ and define $b=\frac{N_b}{N}$. In our case, in one time step $\tau$, only two events are possible -- the number of $\bullet$ particles 
decreases by $2/N$ with probability $\gamma(b)$ or remains constant.

For the inflow dynamics:
\begin{eqnarray}
\gamma^{in}(b) & = b^2,
\end{eqnarray}
and for the outflow dynamics:
\begin{eqnarray}
\gamma^{out}(b) & = (1-b)b^2+W_0b^3=b^2\left[ 1-b(1-W_0) \right].
\end{eqnarray}

It is seen that the above results are not precise, since there is no dependence between $\gamma^{in}(b)$ and $W_0$  for the inflow dynamics and the only stable steady state in this case is $b=0$, i.e. the ferromagnetic state, which is true as long as $W_0>0$. However, as it has been noticed this result is not correct for $W_0=0$.
The same results can be obtained using the mean field approach without mapping. 

However, for the outflow dynamics MFA with dimer mapping gives better results than the basic MFA presented in \cite{SL03}. For $W_0=0$ there are two steady states: $b=0$, i.e. the ferromagnetic state and $b=1$, i.e. the antiferromagnetic state. For $b \ne 0$ and $b \ne 1$,  $\gamma^{out}(b) > 0$ which implies that $b=0$ is an unstable steady state, while $b=1$ is a stable steady state. This result is in agreement with computer simulations \cite{SWS00}.
For $W_0=1$ there is only one ferromagnetic steady state, which was also confirmed by the computer simulations \cite{Krupa}.

The differences between the outflow and inflow dynamics can be already seen if we apply the mean field approach with mapping of the pairs of spins into single particles. In the next section we present simulation results which show yet more differences between these two dynamics.

\section{Simulation results}

The spin updating within both dynamics can be sequential or parallel. In this paper we compare both dynamics for random sequential updating, parallel updating and partially parallel updating. From now on we call the latter case $c$-parallel updating. Within this updating a randomly chosen fraction $c$ of spins is updated synchronously. Of course, $c=1$ corresponds to parallel updating and $c=0$ to random sequential updating.

\subsection{Number of persistent spins}

One of the main quantity of interest in the non-equilibrium dynamics of spin systems at
zero temperature is the fraction of spins, $P(t)$, that persist in the same state up to some
later time $t=N\tau$ (measured in Monte Carlo Steps) \cite{NS99,J02}. In this paper we measure the fraction of persistent spins for both the outflow and inflow dynamics with $c$-parallel updating for different values of $c$. The initial configuration consists of a randomly distributed fraction $p_{+}(0)$ of up spins. 
The number of persistent spins for the outflow dynamics with $W_0=0$ and random sequential updating has been already investigated by Stauffer and Oliveira \cite{SO02} who found agreement with the results for the inflow dynamics, i.e. decays with time $t$ as $1/t^{-3/8}$. However, it was found that in higher dimensions the exponents for the inflow and outflow dynamics are different \cite{SO02}. Here we investigate the case of the Ising spin chain more carefully, i.e. for different values of $W_0$ and $c$.

The first difference between the inflow and outflow dynamics can be already seen for random sequential updating, i.e. $c=0$. For both dynamics the number of persistent spins decays initially as a power-law $\sim t^{-\theta}$. However, for the inflow dynamics the exponent is independent of $W_0$ as long as $W_0>0$, while for the outflow dynamics the exponent is $W_0$-dependent $\theta=\theta(c)$ (see Fig.1). Moreover, for the inflow dynamics the power-law describes properly the decay of the number of persistent spins for all times, while within the outflow dynamics it is valid only for  $t$ smaller than a certain value of time $t^*(W_0)$, dependent on the flipping probability $W_0$. For $W_0 \rightarrow 0$ we obtain $t^*(W_0) \rightarrow \infty$ and the evolution of the number of persistent spins is the same for the outflow and inflow dynamics which is in agreement with the results obtained by Stauffer and Oliveira \cite{SO02}.

\begin{figure}[htb]
\begin{center}
\includegraphics[scale=0.8]{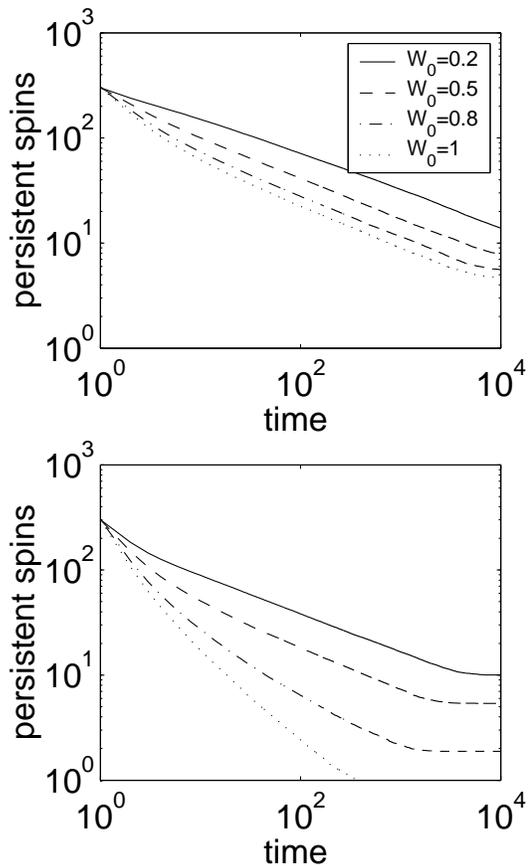}
\caption{The change in time of the number of persistent spins on a chain of $N=300$ spins for random sequential updating (i.e. $c=0$) for the inflow (top) and outflow dynamics (bottom).}
\end{center}
\end{figure}

More differences can be seen for partially synchronous updating with $c>0$. At each elementary time step $\tau$ a fraction $c$ of spins is chosen randomly and is changed synchronously. In such a case we have noticed that the number of persistent spins still decays with a power law for the inflow dynamics. However, for the outflow dynamics the power law is no longer valid. The number of persistent spins decays very fast in this case (see Fig.2). 

\begin{figure}[htb]
\begin{center}
\includegraphics[scale=0.8]{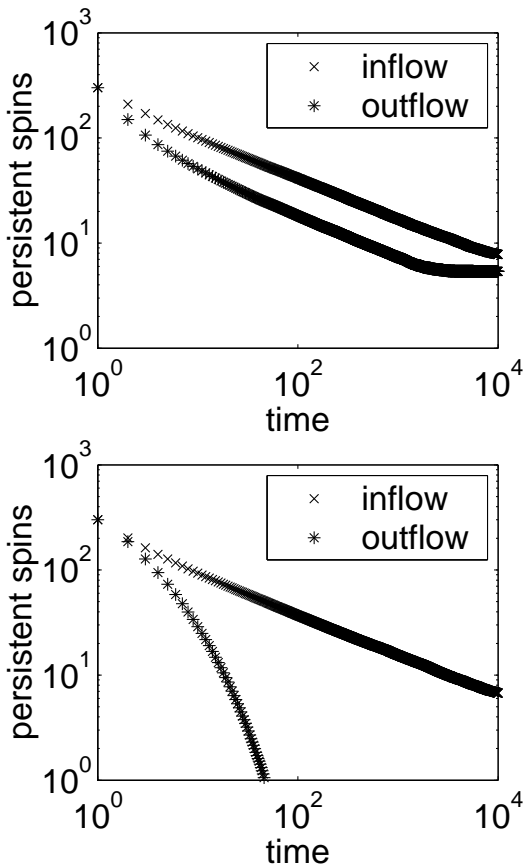}
\caption{The change in time of the number of persistent spins on a chain of $N=300$ spins for partially synchronous updating with $W_0=1/2$. The top panel presents results for $c=0$ and the bottom for $c=0.2$. Obviously for $c>0$ the number of persistent spins decays very fast and cannot be described by a power law.}
\end{center}
\end{figure}

We may conclude this subsection with the following -- the number of persistent spins is $c$ sensible only for the outflow dynamics. For $W_0>0$ and any value of $c$ the number of persistent spins in the inflow dynamics is described by a power law with nearly the same exponent.

\subsection{Number of clusters}

Probably the most natural way to investigate the relaxation process of the consensus dynamics is to look at the number of clusters in time. A cluster consists of a group of spins, each of which is a nearest neighbor to at least one other spin in the cluster, with all spins having the same orientation. With such a definition consensus is reached when only one cluster is present in the system. For both inflow and outflow dynamics with $c$-parallel updating the number of clusters monotonically decays as $t^{-1/2}$ for any value of $c$. This result shows that the number of clusters in time, although a very intuitive and natural measure of  relaxation, is not a good quantity for comparison of the dynamics.

\subsection{Mean relaxation time}

The differences between the dynamics can be observed clearly if we look at the mean relaxation time as a function of the initial fraction of randomly distributed up-spins $p_{+}(0)$. 
Within $0$-parallel updating (i.e. random sequential updating) the relaxation is much slower for the inflow dynamics then for the outflow dynamics. This is also true for the $c$-parallel updating with small $c$. On the contrary, within $1$-parallel updating (i.e. synchronous updating) the relaxation the under outflow dynamics is slower then under the inflow (see Fig.3).  

\begin{figure}
\begin{center}
\includegraphics[scale=0.8]{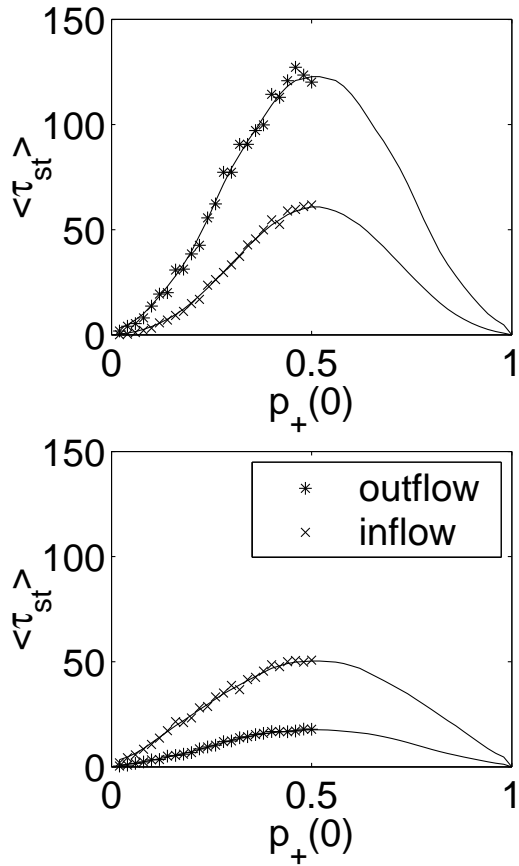}
\caption{The mean relaxation times from the random initial state consisting of $p_{+}(0)$ randomly distributed up spins for $W_0=0.2$. The top panel corresponds to synchronous updating $c=1$, and the bottom panel to $c=0.2$. It can be seen that the relaxation under the outflow dynamics is slower then under the inflow for synchronous updating. On the contrary, the relaxation is much slower under the inflow dynamics then under the outflow dynamics for small $c$.}
\end{center}
\end{figure}

In general, the relaxation times decay with $W_0$ growth, but the dependence between the mean relaxation time and $W_0$ is different for the outflow and inflow dynamics. Two examples for $c=0.2$ and $c=0.5$ for several values of $W_0$ are shown in Figs. 4 and 5, respectively. It can be noted that, for example, for $c=0.5$ and $W_0=0.8$ the dependence between the mean relaxation time and the initial concentration of up-spins $p_{+}(0)$ is nearly the same.  

\begin{figure}
\begin{center}
\includegraphics[scale=0.8]{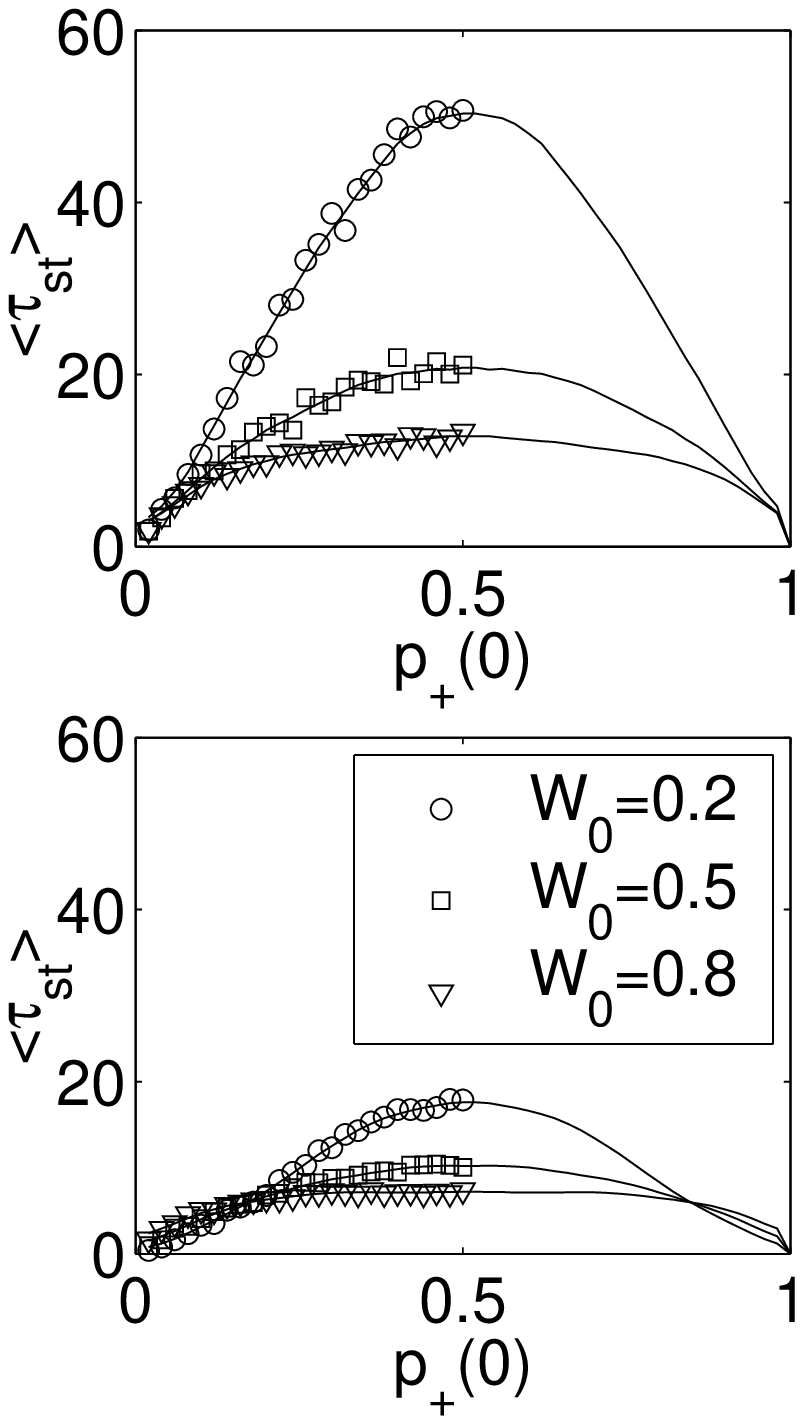}
\caption{The mean relaxation times from the random initial state consisting of $p_{+}(0)$ randomly distributed up spins for $c=0.2$ for the inflow (top) and outflow (bottom) dynamics.}
\end{center}
\end{figure}

\begin{figure}
\begin{center}
\includegraphics[scale=0.8]{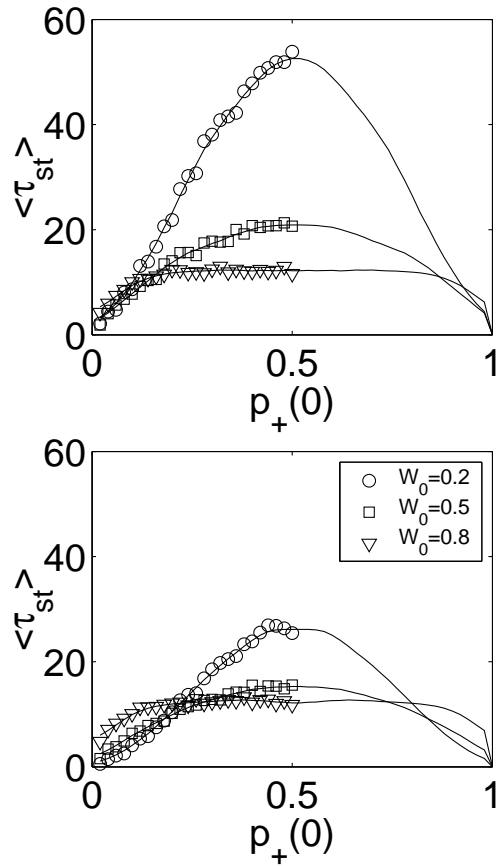}
\caption{The mean relaxation times from the random initial state consisting of $p_{+}(0)$ randomly distributed up spins for $c=0.5$ for the inflow (top) and outflow (bottom) dynamics.}
\end{center}
\end{figure}

In Figure 6 we have presented the dependence between the mean relaxation times from the random initial state consisting of 50\% randomly distributed up spins (maximal waiting time) and the flipping probability $W_0$ for the inflow and outflow dynamics. It is seen that the dependence on $c$ is much stronger for the outflow dynamics. For the inflow dynamics the mean relaxation time is almost the same for all values of $c$. On the other hand for a given value of $c$ the dependence between $<\tau>$ and $W_0$ is stronger for the inflow dynamics.

\begin{figure}
\begin{center}
\includegraphics[scale=0.8]{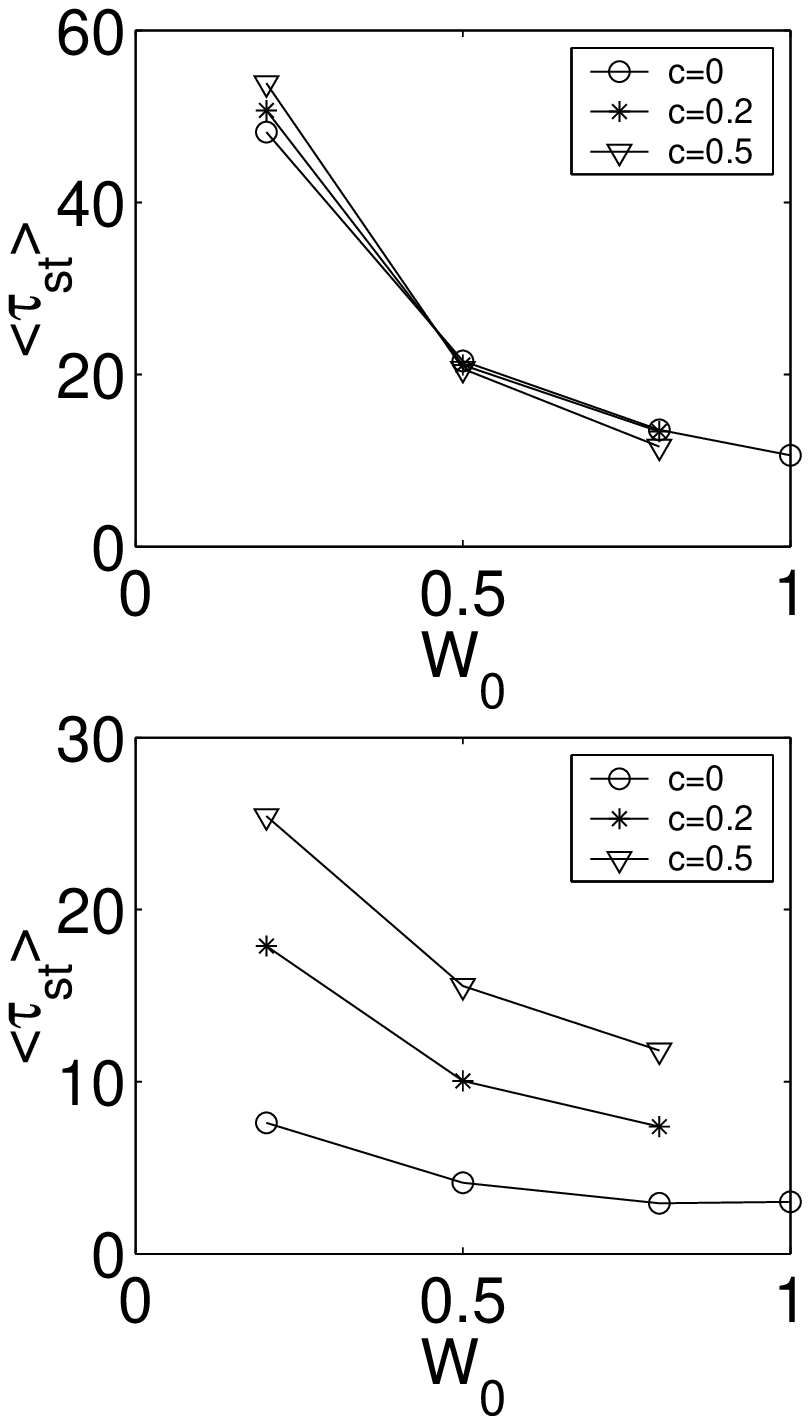}
\caption{The dependence between the mean relaxation times from the random initial state consisting of 50\% randomly distributed up spins and the flipping probability $W_0$ for the inflow (top) and outflow (bottom) dynamics for different values of $c$. It is seen that the dependence on $c$ is much stronger for the outflow dynamics. On the other hand, for a given value of $c$ the dependence between $<\tau>$ and $W_0$ is stronger for the inflow dynamics.}
\end{center}
\end{figure}

\subsection{Distribution of waiting times}
In paper \cite{SL03} the mean field approach for the outflow dynamics with $W_0=0$ has been presented and 
the distribution of waiting times needed to reach the stationary state has been found. Recall that for the $\delta$-initial conditions the distribution of waiting times has an exponential tail \cite{SL03}:
\begin{equation}
P_st^>(\tau) \approx \frac{6}{4}(1-m_0^2)e^{-2\tau}, \tau \rightarrow \infty.
\label{tail}
\end{equation}
Monte Carlo simulations confirm this prediction both on a complete graph and on a chain. In this paper we have checked also the distribution of waiting times for different values of $W_0$ and $c$  both for the outflow and inflow dynamics. It occurs that the distribution of waiting times has an exponential tail for any value of $W_0$ and $c$, although the exponent depends on these parameters. An example for $c=0$, showing the comparison between inflow and outflow dynamics, is shown in Fig. 7.

\begin{figure}
\begin{center}
\includegraphics[scale=0.8]{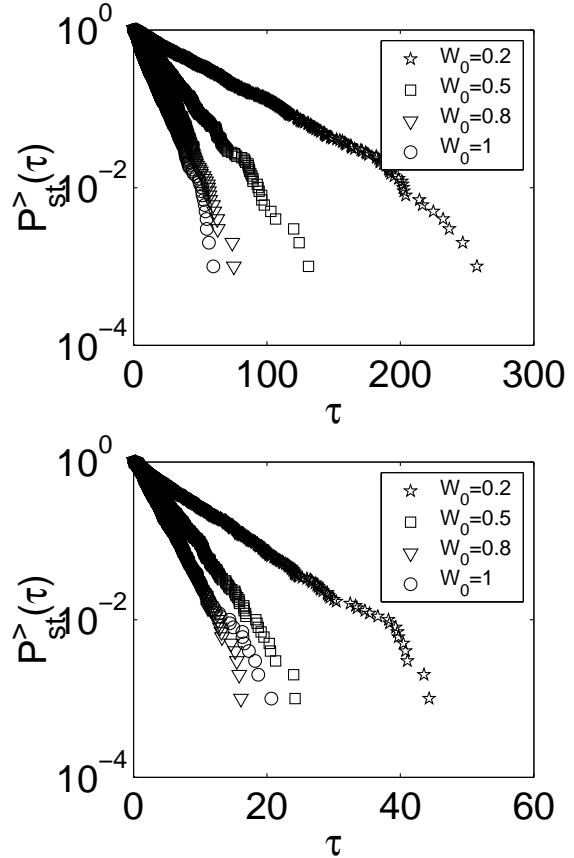}
\caption{Probabilities of reaching the steady state in time
larger than $\tau$ on a chain of $L=200$ spins. 
The distribution of waiting times has an exponential tail for both dynamics -- inflow (top) and outflow (bottom).}
\end{center}
\end{figure}

\section{Conclusions}
It has been suggested \cite{g05, bs03} that the zero-temperature outflow and inflow dynamics for the Ising ferromagnet with the nearest neighbors interactions are equivalent in one dimension. However, this is certainly not true for $W_0 = 0$. This particular value corresponds to the constrained zero-temperature Glauber dynamics where the only possible moves are flips of isolated spins and the system, therefore, eventually reaches a blocked configuration, where there is no isolated spin \cite{GL04}. This can be also easily shown using the method of mapping the Ising spin system onto the RSA dimer model. On the other hand, the outflow dynamics leads to the ferromagnetic steady state for any value of $W_0$. This observation motivated us to compare both dynamics more carefully. We have performed Monte Carlo simulations for both dynamics using random sequential updating, parallel updating and $c$-parallel updating (a randomly chosen fraction $c$ of spins is updated synchronously). We have measured, for different values of $W_0$ and $c$, the distribution of waiting times, the mean waiting time, the decay of the number of clusters and persistent spins in time. 

It occurs that the qualitative difference between the inflow and outflow dynamics is not visible neither in the number of clusters in time nor in the distribution of waiting times. However, it should be noticed that the  relaxation time is different for both dynamics. Nevertheless, for both dynamics the distribution of waiting times has an exponential tail and the number of clusters decays as $t^{-1/2}$ for any value of $W_0>0$ and $c$. 

Differences between the dynamics appear if we look at the dependence between the mean relaxation time and the initial concentration of randomly distributed up spins for different values of $W_0$ and $c$. For $c=0$, which corresponds to  random sequential updating, the mean relaxation time is shorter for the outflow dynamics (e.g. for $W_0=0.2$ and $p_0=0.5$ it is about 10 times shorter) than for the inflow. The mean relaxation time $<\tau>$ decreases with $W_0$ increase for both dynamics, but the dependence between $<\tau>$ and $W_0$ is different for the outflow and inflow dynamics. Generally, the mean relaxation time decays faster with increase $W_0$ for the inflow dynamics for any value of $c$. Moreover, with increase $c$ the dependence between the mean relaxation time for the inflow dynamics and the outflow dynamics vanishes. As the results for some values of $c$ and $W_0$ (e.g. $c=0.5$ and $W_0=0.8$) the dependence between the mean relaxation times and the initial concentration of up spins is identical. Of course, this suggests that for some values of parameters $W_0$ and $c$ the relaxation under outflow dynamics is faster than under the inflow dynamics. This is indeed true. For $c=1$ (parallel updating), the relaxation is faster under the inflow dynamics for any value of $W_0$.

The second quantity which occurs to behave differently for both dynamics is the number of persistent spins in time. Main differences are seen for partially synchronous updating with $c>0$. In such a case we have noticed that the number of persistent spins decays with a power law for the inflow dynamics (like for $c=0$). However, for the outflow dynamics the power law is no longer valid. The number of persistent spins decays very fast in this case.

Concluding, the inflow and outflow dynamics differ very clearly even in one dimension. There is obvious, very strong difference for $W_0=0$, but also for $W_0>0$ both dynamics are qualitatively different. In the case of random sequential updating the relaxation under the outflow dynamics is much faster than under the inflow dynamics. On the contrary, in the case of parallel updating the relaxation for the outflow dynamics is much slower than for the inflow.
It should be mentioned here that the outflow dynamics with $W_0=0$ and synchronous updating has been investigated earlier and it has been found that in such a case the possibility of reaching a consensus is reduced quite dramatically \cite{SR03}. Also the number of persistent spins is $c$ sensible only for the outflow dynamics. For $W_0>0$ and any value of $c$ the number of persistent spins in the inflow dynamics is described by a power law with nearly the same exponent.
Generally, it occurs that the outflow dynamics is much more influenced by the type of updating than the inflow dynamics. 
We believe that this result is very important in the various interdisciplinary applications of the zero-temperature single-spin flip dynamics.

\end{document}